\documentclass[3p]{elsarticle}

\usepackage{graphicx}
\usepackage{titlesec}

\usepackage[T1]{fontenc}
\usepackage[latin9]{inputenc}
\usepackage{amssymb}
\usepackage{amsmath}
\usepackage{multicol}
\usepackage{amsthm}
\usepackage[figuresright]{rotating}
\usepackage{bm}
\usepackage{caption}

\usepackage{lineno}

\usepackage{xcolor}
\usepackage{pdfcolmk}
\usepackage{textcomp}
\usepackage{stackrel}
\usepackage{graphicx}
\usepackage{refcount}
\usepackage{esint}
\usepackage{hhline}
\usepackage{booktabs}
\usepackage{xfrac}

\usepackage{booktabs}
\PassOptionsToPackage{normalem}{ulem}
\usepackage{ulem}
\usepackage[unicode=true,
 bookmarks=true,bookmarksnumbered=true,bookmarksopen=true,bookmarksopenlevel=1,
 breaklinks=false,pdfborder={0 0 0},pdfborderstyle={},backref=false,colorlinks=false]
 {hyperref}
\hypersetup{pdftitle={Your Title},
 pdfauthor={Your Name},
 pdfpagelayout=OneColumn, pdfnewwindow=true, pdfstartview=XYZ, plainpages=false}

\PassOptionsToPackage{unicode}{hyperref}
\PassOptionsToPackage{naturalnames}{hyperref}
\usepackage[cmintegrals]{newtxmath}
\usepackage{bm}
\usepackage{algorithmic}
\usepackage{xfrac}
\usepackage{array}
\usepackage{makecell}
\usepackage[caption=false,font=normalsize,labelfont=sf,textfont=sf]{subfig}
\usepackage{float}
\usepackage{dblfloatfix}
\usepackage{cases}
\usepackage{url}
\usepackage[caption=false,font=normalsize,labelfont=sf,textfont=sf]{subfig}
\newtheorem{thm}{Theorem}
\newtheorem{lem}[thm]{Lemma}
\newtheorem{cor}{Corollary}[thm]
\newdefinition{rmk}{Remark}
\newproof{pf}{Proof}
\newproof{pot}{Proof of Theorem \ref{thm2}}

\makeatother

\usepackage{fancyhdr}

\fancyheadoffset[RO,EL]{0pt}
\usepackage{geometry}

\begin{document}
\begin{frontmatter}

\title{
\begin{center}
{\bf \ Capacity Analysis of the Fluctuating Double-Rayleigh with Line-of-Sight Fading Channel\tnoteref{t1}}
\tnotetext[t1]{\copyright 2022. This manuscript version is made available under the CC-BY-NC-ND 4.0 (\href{license https://creativecommons.org/licenses/by-nc-nd/4.0/}{license https://creativecommons.org/licenses/by-nc-nd/4.0/})}
\end{center}
}
 %

\author[]{Aleksey S. Gvozdarev}

\address{P. G. Demidov Yaroslavl State University
}

\cortext[]{Aleksey S. Gvozdarev (asg.rus@gmail.com)}
\cortext[]{This work was supported by Russian Science Foundation under Grant 22-29-01458 (https://rscf.ru/en/project/22-29-01458/).}

\begin{abstract}

The proposed research performs  the capacity analysis of the wireless channel described by the fluctuating double-Rayleigh with the line-of-sight model. The closed-form analytical expressions  for the conditional capacity (in the case of arbitrary noninteger fading parameter) and the ergodic capacity (in the case of the integer fading parameter) are derived in terms of the extended generalized bivariate Meijer G-function. The exact solutions are succeeded by the approximating expressions deduced for the cases of small and large ratios between the Rician K-factor and the fading parameter. The performed numeric simulation verifies the correctness of the derived results and analyzes the proposed capacity approximation quality performance.

\end{abstract}


\end{frontmatter}

\section{Introduction}
Nowadays, the capacity of ad hoc wireless communication systems is mainly limited by the effects in the propagation channel. In the situation of the increase of the operating frequency as well as the frequency range, importance is being increasingly attached to such factors as multipath fading and shadowing. Although the existence of the direct path between the transmitter and the receiver (i.e. line-of-sight (LoS)) is not a prerequisite condition for wireless communications, in many modern applications the shrinkage of the coverage area leads to its emergence.

Most of the contemporary fading models that assume LoS propagation \cite{Sha10} are built upon the framework proposed by Rice \cite{Ric44}. To broaden the application, incorporating a greater number of propagation factors (reflection, diffraction, multipath etc.), it was further generalized to account for multiple scattering \cite{And02, Sal05} (for example, second-order scattering fading \cite{Lop18, Lop20}), leading to the so-called double-Rayleigh and double-Rayleigh with LoS models \cite{Sal06, Pan18, Pan20}, including such cases as the keyhole created by the "pipe-like" channel \cite{Ges02}, amplify-and-forward relay \cite{Has04}, and propagation via diffracting street corner \cite{Erc97}. Such a model has been also extensively used in free-space optical communication through turbulent medium \cite{And85} (the so-called IK model), and Vehicle-to-vehicle (V2V) communications \cite{Ai18}.

An alternative approach extends the classical model by accounting for possible effects of LoS component shadowing leading to the Rician shadowed class of models that differ by the way shadowing effects are introduced (see, for instance, \cite{Abd03, Par10}).

The recent achievements in this area are mainly associated with the combination of the aforementioned approaches as it was proposed in \cite{Lop21}, where the double-Rayleigh model \cite{Sal06} was further generalized by introducing the Gamma-distributed shadowing of the LoS component. It was demonstrated (see Section III in \cite{Lop21}) that such a model helps to avoid the overestimation of the peak probability density existing in the classical double-Rayleigh with LoS models (dRLoS) (see \cite{Chu89}, for instance, for the IK model).

For such a model, the closed-form expression for the probability density function (pdf) was obtained in \cite{Lop21}, although in the form that does not make further analytical derivations possible. For a specific case of the integer fading parameter, a simplified expressions for the pdf and cumulative density function were obtained and applied to the problem of the outage probability analysis \cite{Lop21}. Later, a closed-form bit error rate performance was examined, and the impact of fading parameters on the error rate was discussed \cite{Gvo22a}. Nonetheless, the capacity study is absent.

Motivated by the model proposed in  \cite{Lop21}, the present research fills the existing gap in the system performance analysis and derives the capacity of the wireless channel described by the fluctuating double-Rayleigh with the line-of-sight model (fdRLoS).  To achieve this objective, several closed-form analytical results were derived and analyzed.
The major contributions of this work can be summarized as follows:
\begin{itemize}
  \item For the optimal rate adaptation strategy with constant transmit power (ORA) following closed-form expressions are derived: \textit{a)}~the conditional capacity for the case of arbitrary noninteger fading parameter, \textit{b)}~an expression for the ergodic capacity in case of the integer fading parameter, \textit{c)}~computationally efficient analytic approximations deduced for the cases of small and large ratios between the Rician K-factor and the fading parameter, \textit{d)}~an asymptotic expression for the ergodic capacity in the case of high signal-to-noise ratio.
   \item For the optimal simultaneous power and rate adaptation strategy (OPRA) following closed-form expressions are derived: \textit{a)}~an expression for the ergodic capacity in case of arbitrary fading parameter, \textit{b)}~an asymptotic expression for the ergodic capacity in the case of high signal-to-noise ratio.
  \item A connection between the ORA and OPRA ergodic capacity expressions in the case of high signal-to-noise ratio is established.
\end{itemize}

Lastly, to validate the accuracy of the analytical work, computer simulation and numeric analysis were performed, and the obtained results concerning the impact of fading parameters on the proposed capacity approximation of the fdRLoS channel were studied.

\section{Problem description}
\subsection{Channel model: physical and statistical description}
The proposed in \cite{Lop21} fluctuating double-Rayleigh with line-of-sight fading channel model assumes that the received signal $S$ can be represented as the linear combination of the shadowed LoS component (with average magnitude $\omega_0$ and uniformly distributed phase $\phi                          \sim U[0,2\pi )$) and double-Rayleigh fading (dRf) component $\omega_2 G_2 G_3$:
\begin{equation}\label{channel-model}
S=\omega_0\sqrt{\xi }e^{j\phi }+\omega_2 G_2 G_3,
\end{equation}
where $\xi$ is the Gamma distributed shadowing parameter with unit power shape coefficient $m$, $\omega_2$ is the average dRf component magnitude, and $G_2, G_3$ are the zero-mean complex normal random variables (i.e., $G_2, G_3                          \sim \mathcal{CN}(0,1)$).

Physically this model corresponds to the propagation channel with constant amplitude LoS component (first term in \eqref{channel-model}) that  undergoes shadowing with unit power and multipath fading components (second term) that are subjected to second-order scattering.

Following \cite{Lop21}, let us assume that the channel is properly normalized to yield unit power of the received signal (i.e. $\mathbb{E}{|S|^2}=1$), and the probability density function (pdf) of the instantaneous signal-to-noise ratio $\gamma$, defined as $\gamma=\bar{\gamma }|S|^2$ (with $\bar{\gamma }$ -- the average signal-to-noise ratio) is given by (see equation 13 in \cite{Lop21})
\begin{equation} \label{eq-pdf-full}
f_{\gamma }(\gamma )=\int_{0}^{\infty}f_{\gamma_x}(\gamma|x) e^{-x}{\rm d}x,
\end{equation}
where $f_{\gamma_x}(\gamma|x)$ denotes the conditional probability density function conditioned on $x=|G_3|^2$, which is defined as
\begin{eqnarray} \label{eq-pdf-conditional}
f_{\gamma_x}(\gamma|x)=\frac{m^m(1+k_x)}{(m+k_x)^m\bar{\gamma }_x}e^{-\frac{1+k_x}{\bar{\gamma }_x}\gamma }  \mbox{}_1F_1\left(m,1,\frac{k_x(1+k_x)}{k_x+m}\frac{\gamma }{\bar{\gamma }_x}\right),
\end{eqnarray}
with $\mbox{}_1F_1(                         \cdot )$ being the confluent hypergeometric function \cite{DLMF}, $\bar{\gamma }_x=\frac{k+x}{k+1}\bar{\gamma }$, $k_x=\frac{k}{x}$, $k=\frac{\omega_0^2}{\omega^2_2}$ -- the Rician K-factor.

It can be noticed that the assumed channel (proposed in \cite{Lop21}) closely resembles the double Rician model widely used nowadays for the analysis of the reconfigurable intelligent surfaces' assisted communications (see, for instance, \cite{Tao20, Che22}). The main problem with the double Rician model is that due to its complexity there is no tractable expression for the pdf of the instantaneous SNR (which is present for the fdRLoS channel), moreover, it does not handle the possible shadowing of the LoS components.

\subsection{System performance metric}
The research assumes an ergodic capacity $\bar{\bm{\mathrm{C}}}$ \cite{Big98} as the primary metric that is widely used to quantify the communication link quality. The exact expression for the capacity depends on the employed adaptive transmission schemes, which rely upon the available channel state information (SCI). Herein one assumes two of the most popular transmission strategies \cite{Pep11}: optimal rate adaptation with constant transmit power (ORA), applied for the cases when CSI is available at the receiver only; and optimal simultaneous power and rate adaptation (OPRA), when CSI is available at the transmitter.

For a multipath fading channel $\bar{\bm{\mathrm{C}}}$ is defined as a system capacity \cite{Gol97} (per unit bandwidth) averaged over the stochastic variations of the instantaneous signal-to-noise ratio~ \cite{Pep11}
\begin{equation} \label{eq-cap-def-1-ORA}
\mathbf{ORA}: \bar{\bm{\mathrm{C}}}=\int_{0}^{\infty}\log_2(1+\gamma )f_{\gamma }(\gamma ){\rm d}\gamma,
\end{equation}

\begin{align} \label{eq-cap-def-1-OPRA}
\mathbf{OPRA}:\begin{cases}
       & \bar{\bm{\mathrm{C}}}=\displaystyle\int_{\gamma_0}^{\infty}\log_2\left(\frac{\gamma }{\gamma_0} \right)f_{\gamma }(\gamma ){\rm d}\gamma ,\\[4ex]
      \gamma_0: & \displaystyle\int_{\gamma_0}^{\infty}\left(\frac{1}{\gamma_0}-\frac{1}{\gamma } \right)f_{\gamma }(\gamma ){\rm d}\gamma=1,
    \end{cases}
\end{align}
where $\gamma_0$ is the optimal cut-off threshold SNR.

For the assumed problem the pdf $f_{\gamma }(\gamma )$ does not allow direct evaluation of \eqref{eq-cap-def-1-ORA} or \eqref{eq-cap-def-1-OPRA} without specific simplifications. But the problem can be resolved by using a following two-step procedure: 1) evaluate \eqref{eq-cap-def-1-ORA} or \eqref{eq-cap-def-1-OPRA} with pdf $f_{\gamma_x}(\gamma|x)$, thus obtaining the conditional ergodic capacity $\bar{\bm{\mathrm{C}}_x}$, 2) perform the Laplace transform (which is equivalent to averaging the result with the exponential pdf, see \cite{Lop21}), i.e.
\begin{align} \label{eq-cap-def-2}
 \bar{\bm{\mathrm{C}}}=\begin{cases}
       \int_{0}^{\infty}\underbrace{\left\{\int_{0}^{\infty}\log_2(1+\gamma )f_{\gamma_x}(\gamma|x){\rm d}\gamma \right\}}_{\bar{\bm{\mathrm{C}}_x}} e^{-x}{\rm d}x & \text{for}\quad \mathbf{ORA}, \\[4ex]
      \int_{0}^{\infty}\underbrace{\left\{\int_{\gamma_0}^{\infty}\log_2\left(\frac{\gamma }{\gamma_0} \right)f_{\gamma_x}(\gamma|x){\rm d}\gamma \right\}}_{\bar{\bm{\mathrm{C}}_x}} e^{-x}{\rm d}x & \text{for}\quad \mathbf{OPRA}.
    \end{cases}
\end{align}

Hence in the following section  \eqref{eq-cap-def-2} will be used to evaluate~$\bar{\bm{\mathrm{C}}}$.

\section{Performance analysis}
Since further analysis depends on which adaptation scheme one uses, one performs sequential derivations first for ORA strategy and then for OPRA.

\subsection{ORA case}
For the fdRLoS channel, it should be noted that $f_{\gamma }(\gamma )$ has two distinct forms (see equations (5) and (7) in \cite{Lop21}) depending on $m$: $m$ is noninteger and $m$ is an integer.

In the first case, $\bar{\bm{\mathrm{C}}}_x$ is given by the following Lemma.
\begin{lem}
The conditional ergodic capacity for the fluctuating double-Rayleigh with line-of-sight fading channel model (with $m                       \notin                       \mathbb{Z}^{+}_0$) in case of the optimal rate adaptation strategy can be represented as
\begin{eqnarray}\label{lem-1}
\bar{\bm{\mathrm{C}}}_x=\left(\frac{m x}{k+m x}\right)^{m-1}\frac{1}{\Gamma (1-m)\ln 2 }         G_{1,0:2,2:1,2}^{0,1:2,0:1,1}\left(
\begin{array}{c}
0\\
\mbox{---}\\
\end{array}\middle\vert
\begin{array}{c}
1, 1\\
1, 0\\
\end{array}\middle\vert
\begin{array}{c}
m \\
0, 0\\
\end{array}\middle\vert
\frac{\bar{\gamma }(k+m x)}{m(k+1)},\frac{k}{m x}
\right),
\end{eqnarray}
where $G_{1,0:2,2:1,2}^{0,1:2,0:1,1}$ is the extended generalized bivariate Meijer G-function (EGBMG) (see \eqref{EGNMGF} at the top of the next page, with the integration contours $\mathcal{L}_s, \mathcal{L}_t$ defined as in \cite{Hai92}), and $\Gamma (                  \cdot )$ is the Euler gamma-function.
\end{lem}
\begin{pf}

\begin{figure*}[!t]
\normalsize
\vspace*{-4pt}
\begin{eqnarray}\label{EGNMGF}
&&\hspace{-40pt}G_{q_1,p_1:p_2,q_2:p_3,q_3}^{n_1,m_1:m_2,n_2:m_3,n_3}\left(
\begin{array}{c}
\left(a_{p_1}\right)\\
\left(b_{p_1}\right)
\end{array}\middle\vert
\begin{array}{c}
\left(a_{p_2}\right)\\
\left(b_{p_2}\right)
\end{array}\middle\vert \;
\begin{array}{c}
\left(a_{p_3}\right)\\
\left(b_{p_3}\right)
\end{array}\middle\vert
x,y
\right)=\frac{1}{(2\pi j)^2}\int_{\mathcal{L}_s}\int_{\mathcal{L}_t}\Psi_1(s+t)\Psi_2(s)\Psi_3(t) x^{-s}y^{-t}\;{\rm d}s\;{\rm d}t  \\
&&\text{with \qquad} \Psi_k(\tau )=\frac{\prod                           _{j=1}^{m_k}\Gamma (b^{k}_j+\tau )\prod                           _{j=1}^{n_k}\Gamma (1-a^{k}_j-\tau )}
{\prod                           _{j=n_k+1}^{p_k}\Gamma (a^{k}_j+\tau )\prod                           _{j=m_k+1}^{q_k}\Gamma (1-b^{k}_j-\tau )}, \qquad k=1,2,3. \nonumber
\end{eqnarray}
\hrulefill
\end{figure*}

To prove Lemma 1, one starts with the definition of the conditional ergodic capacity $\bar{\bm{\mathrm{C}}}_x$ in \eqref{eq-cap-def-2}.
Applying the relations between the integrands and Meijer G-functions (see \cite{Pru92} 8.4.6.5, 8.4.3.1 and 8.4.45.5 respectively):
\begin{eqnarray}
&&\ln(1+\gamma )=G_{2,2}^{\,1,2}\!\left(\left.{\begin{matrix}1,1\\1,0\end{matrix}}\;\right|\,\gamma \right),   \label{eq_lem-1-1}        \\
&& e^{-\frac{(k+1)}{x\bar{\gamma }}\gamma }=G_{0,1}^{\,1,0}\!\left(\left.{\begin{matrix}\mbox{---}\\0\end{matrix}}\;\right|\,\frac{(k+1)\gamma }{x\bar{\gamma }}\right),  \label{eq_lem-1-2}         \\
&&\hspace{-30pt}\mbox{}_1F_1\left(m,1,\frac{k(k+1)\gamma }{x(k+m x)\bar{\gamma }}\right)=\frac{e^{\frac{k(k+1)\gamma }{x(k+m x)\bar{\gamma }}}}{\Gamma (1-m)}  G_{1,2}^{\,1,1}\!\left(\left.{\begin{matrix}m\\0,0\end{matrix}}\;\right|\,\frac{k(k+1)\gamma }{x(k+m x)\bar{\gamma }}\right),\label{eq_lem-1-3}
\end{eqnarray}
and making use of the Meijer G-function integration theorem (see \cite{Wol21}) finalizes the proof.
\end{pf}

Combining the definition \eqref{eq-cap-def-1-ORA} and Lemma 1 yields the capacity for the assumed channel model. The resultant integral can be efficiently evaluated numerically, but up to now no closed-form analytical solution exists. Although the EGBMG function (contrary to the classical Meijer G-function) is not readily implemented in modern computer algebra systems, it can be computed via the procedures presented in \cite{Ans11} or \cite{Gar14}.

It should also be noted that the solution is valid for $m                       \notin                       \mathbb{Z}^{+}_0$ since EGBMG is defined in terms of the double Mellin-Barnes integral with the integration contours in the corresponding complex domains chosen in such a way to separate the specific singularities of the integrand, but for $m                       \in \mathbb{Z}^{+}_0$ no such contour can be found. Moreover, EGBMG in \eqref{lem-1} alternates the sign with the increase of $m$, the same goes for $\Gamma (1-m)$ thus making $\bar{\bm{\mathrm{C}}}_x$ nonnegative.

However,  in \cite{Lop21} it was demonstrated that, from a practical perspective, $m$ can be efficiently treated as an integer. In this case, the expression for the ergodic capacity \eqref{eq-cap-def-1-ORA} can be obtained in the closed-form, as stated by the following theorem.
\begin{thm}
The ergodic capacity expression for fluctuating double-Rayleigh with line-of-sight fading channel model with ($m                          \in \mathbb{Z}^{+}$) in case of the optimal rate adaptation strategy can be represented as
\begin{eqnarray}\label{thm-1}
\bar{\bm{\mathrm{C}}}=\left(\frac{\bar{\gamma }}{k+1}\right)^{m-1}\frac{\Gamma (m)}{\ln 2 }\sum_{i=0}^{m-1}\frac{\left(\frac{k}{m}\right)^i}{(i!)^2\Gamma (m-i)}   G_{1,1:0,1:2,3}^{1,1:1,0:3,1}\left(
\hspace{-5pt} \begin{array}{c}
m-1, m\\
m-1, m-1, m+i\\
\end{array}\hspace{-2pt}\middle\vert
\begin{array}{c}
\mbox{---}\\
0\\
\end{array}\hspace{-2pt}\middle\vert
\begin{array}{c}
1 \\
i\\
\end{array}\middle\vert
\frac{k}{m},\frac{k+1}{\bar{\gamma }}
\right).
\end{eqnarray}
\end{thm}
\begin{pf}
To prove Theorem 2, one can note that the probability density function for the fdRLoS model \eqref{eq-pdf-conditional} coincides with the probability density function for
the $\kappa-\mu$ shadowed channel model (see \cite{Gar14}, equation 2) up to the substitution $\mu=1$.
Thus, the conditional ergodic capacity in the case of $m                          \in \mathbb{Z}^{+}$ can be derived as
\begin{eqnarray}\label{eq_th-1-2}
\bar{\bm{\mathrm{C}}}_x=\frac{(1)_{m-1}\left(\frac{x m}{k+m x}\right)^{m-1}}{\Gamma (m)\ln 2}\sum_{i=0}^{m-1}\frac{(1-m)_{i}}{(i!)^2}                   \left(-\frac{k}{m x}\right)^i G_{3,2}^{\,1,3}\!\left(\left.{\begin{matrix}1,1,-i\\1,0\end{matrix}}\;\right|\,\frac{\bar{\gamma }(k+m x)}{m(k+1)}\right).
\end{eqnarray}

Combining  \eqref{eq_th-1-2} and \eqref{eq-cap-def-2} yields the following:
\begin{eqnarray}\label{eq_th-1-3}
\bar{\bm{\mathrm{C}}}=\left(\frac{\bar{\gamma }}{k+1}\right)^{m-1}\frac{1}{\ln 2}\sum_{i=0}^{m-1}\frac{(1-m)_{i}}{(i!)^2}\left(-\frac{k}{m}\right)^i  \displaystyle \int_{0}^{\infty}x^{m-i-1} G_{3,2}^{\,1,3}\!\left(\left.{\begin{matrix}2-m,2-m,1-m-i\\2-m,1-m\end{matrix}}\;\right|\,\frac{\bar{\gamma }(k+m x)}{m(k+1)}\right)e^{-x}{\rm d}x,
\end{eqnarray}
where the following property of the Meijer G-function was used (see equality 16.19.2 in \cite{DLMF}):
\begin{equation}\label{eq_th-1-4}
z^{\alpha }G_{p,q}^{\,m,n}\!\left(\left.{\begin{matrix}(a_p)\\(b_q)\end{matrix}}\;\right|\,z\right) =G_{p,q}^{\,m,n}\!\left(\left.{\begin{matrix}(a_p)+\alpha \\(b_q)+\alpha \end{matrix}}\;\right|\,z\right).
\end{equation}

The Meijer G-function can be represented via a contour integral in complex domain (see definition 16.17.1 in \cite{DLMF})
\begin{eqnarray}\label{eq_th-1-5}
G_{3,2}^{\,1,3}\!\left(\!\left.{\begin{matrix}2-m,2-m,1-m-i\\2-m,1-m\end{matrix}}\;\right|\,\frac{\bar{\gamma }(k+m x)}{m(k+1)}\right)=\frac{1}{2 \pi  j}\int _{\mathcal{L}_s}\!\!\frac{\Gamma (2-m-s)\Gamma (m+i+s)}{ \Gamma (m+s)\Gamma^{-2} (m+s-1)}\!\left(\frac{\bar{\gamma }(k+m x)}{m(k+1)}\right)^s \!\!{\rm d}s
\end{eqnarray}
with the contour  $\mathcal{L}_s$ satisfying the convergence of the integral (i.e. defined as in  \cite{DLMF} (see 16.17.1, case "i")).

Combining \eqref{eq_th-1-5} and \eqref{eq_th-1-3}, switching the order of integration  and noting that the integral over the variable $x$ evaluates to the confluent Kummer hypergeometric function $U(                           \cdot )$, i.e.
\begin{eqnarray}\label{eq_th-1-6}
\int_{0}^{\infty}\!\!\! x^{m-i-1} (k+m x)^s e^{-x}{\rm d}x= m^{i-m}\Gamma (m-i)   k^{m-i-1}  U\left(m-i,m-i+s+1,\frac{k}{m}\right),
\end{eqnarray}
and applying Kummer's transformations (see definition 13.2.40 in \cite{DLMF}) one can get the following representation for $\bar{\bm{\mathrm{C}}}$:
\begin{eqnarray}\label{eq_th-1-7}
\bar{\bm{\mathrm{C}}}=\frac{\left(\frac{\bar{\gamma }}{k+1}\right)^{m-1}\Gamma (m)}{2\pi j\ln 2 }\sum_{i=0}^{m-1}\frac{\left(\frac{k}{m}\right)^i}{(i!)^2}  \int_{\mathcal{L}_s}\frac{\Gamma (2-m-s)}{ \Gamma (m+s)}        \frac{\Gamma (m+i+s)}{\Gamma^{-2} (m+s-1)} U\left(-s,i+1-m-s,\frac{k}{m}\right)\left(\frac{\bar{\gamma }}{k+1}\right)^{s} {\rm d}s.
\end{eqnarray}

Using the Mellin-Barnes transform for $U(                           \cdot )$, i.e.
\begin{eqnarray}\label{eq_th-1-8}
U\left(-s,i+1-m-s,\frac{k}{m}\right)=\frac{1}{2\pi j \Gamma (-s)\Gamma (m-i)}                  \int_{\sigma-j\infty}^{\sigma+j\infty}\Gamma (t)\Gamma (-s-t)\Gamma (m-i+s+t)
\left(\frac{m}{k}\right)^{t} {\rm d}t
\end{eqnarray}
with $\sigma$ satisfying the integral convergence region. Applying the definition of the extended generalized bivariate Meijer G-function (see \eqref{EGNMGF}) with the proper choice of the integration contours $\mathcal{L}_s, \mathcal{L}_t$ (defined, for instance, as in \cite{Hai92}) completes the proof.
\end{pf}

Although the derived expression can be successfully evaluated via numeric computational modules for EGBMG  calculation (see \cite{Ans11} or \cite{Gar14}), the procedure is quite complex: special care is needed to choose the integration contours that guarantee the convergence (they will heavily depend upon the channel parameters), and the double numerical integration process is computationally expensive.

Thus, from a practical point of view,  it would be valuable to have at hand a computationally efficient approximation of \eqref{thm-1}. Since we are mainly interested in the expression valid for any possible varying SNR for a given channel (i.e. fixed channel parameters), it is useful to assume the asymptotical behavior of the integrand depending on the ratio $\sfrac{k}{m}$.
\begin{thm}
The ergodic capacity of the fluctuating double-Rayleigh with line-of-sight fading channel model with ($m                          \in \mathbb{Z}^{+}$) in case of the optimal rate adaptation strategy can be approximated as
\begin{itemize}
  \item in cases of small Rician factor $k$ (with moderate values of shadowing factor $m\approx 1$) or high $m$ (with moderate~$k$)
  \begin{eqnarray}\label{thm-2-1}
\bar{\bm{\mathrm{C}}}_{\text{ap}}^{\sfrac{k}{m}-\text{low}}         \simeq         \frac{1}{\ln 2 }G_{4,2}^{\,1,4}\!\left(\left.{\begin{matrix}1,1,0,0\\1,0\end{matrix}}\;\right|\,\frac{\bar{\gamma }}{k+1}\right)+ \frac{k(m-1)}{m\ln 2}G_{4,2}^{\,1,4}\!\left(\left.{\begin{matrix}1,1,-1,1\\1,0\end{matrix}}\;\right|\,\frac{\bar{\gamma }}{k+1}\right),
\end{eqnarray}
  \item  in cases of high Rician factor $k$ (with moderate values of shadowing factor $m$):
  \begin{eqnarray}\label{thm-2-2-1}
\bar{\bm{\mathrm{C}}}_{\text{ap}1}^{\sfrac{k}{m}-\text{hi}}                           \simeq                           \frac{\Gamma (m)}{\ln 2 }\sum_{i=0}^{m-1}\sum_{l=0}^{n-1}\frac{(-1)^l\left(\frac{k}{m}\right)^{i+1-m-l}(m-i)_l}{(i!)^2l!}           G_{4,3}^{\,2,3}\!\left(\left.{\begin{matrix}1,1,-i,m-1\\1, l+m-1, 0\end{matrix}}\;\right|\,\frac{k\bar{\gamma }}{m(k+1)}\right),
\end{eqnarray}
  \begin{eqnarray}\label{thm-2-2-2}
\bar{\bm{\mathrm{C}}}_{\text{ap}2}^{\sfrac{k}{m}-\text{hi}}                           \simeq                           \frac{k \Gamma (m)}{m\ln 2 }\sum_{i=0}^{m-1}\frac{\left(\frac{m}{k+m}\right)^{m-i}}{(i!)^2} G_{4,3}^{\,2,3}\!\left(\left.{\begin{matrix}1,1,-i,m-1\\1, m-1, 0\end{matrix}}\;\right|\,\frac{k\bar{\gamma }}{m(k+1)}\right),
\end{eqnarray}
\end{itemize}
where  $(                  \cdot )_l$ is the Pochhammer symbol and $n$ is the number of terms, defining the approximation quality.
\end{thm}
\begin{pf}
To prove Theorem 3, one starts with the expression \eqref{eq_th-1-7}. It is obvious that the summands depend on the ratio $\frac{k}{m}$, and thus an asymptotic series expansion of  $U(        \cdot )$ for the cases $\frac{k}{m}\to 0$ and $\frac{k}{m}\to \infty$ can be applied.

In the first case (i.e., $\frac{k}{m}\to 0$),
\begin{equation}\label{eq_th-3-0}
 U\left(-s,i+1-m-s,\frac{k}{m}\right)\bigg|_{\frac{k}{m}\to 0}       \sim \frac{\Gamma (m+s-i)}{\Gamma (m-i)}.
\end{equation}

Thus, the expression for $\bar{\bm{\mathrm{C}}}_{\text{ap}}^{\text{low}}$ turns into
\begin{equation}\label{eq_th-3-1}
\bar{\bm{\mathrm{C}}}_{\text{ap}}^{\sfrac{k}{m}-\text{low}}=\frac{\left(\frac{\bar{\gamma }}{k+1}\right)^{m-1}\Gamma (m)}{\ln 2 }\sum_{i=0}^{m-1}\frac{\left(\frac{k}{m}\right)^i}{(i!)^2\Gamma (m-i)}\frac{1}{2\pi j}   \int_{\mathcal{L}_s}\frac{\Gamma (2-m-s)\Gamma (m+i+s)}{ \Gamma (m+s)\Gamma^{-2}(m+s-1)}\Gamma (m-i+s) \left(\frac{\bar{\gamma }}{k+1}\right)^{s} {\rm d}s.
\end{equation}

It should be specifically pointed out that the contour-integral represents Meijer G-function only in the case,
when the singularities of the numerator on the positive (with $+s$) and negative (with $-s$) sides of the real axis can be separated.
But this is possible only when $m<2$, and since $m                          \in \mathbb{Z}^{+}$, the sum degenerates into two terms, thus completing the first part of the proof.

In the second case,
\begin{equation}
 U\left(-s,i+1-m-s,\frac{k}{m}\right)\bigg|_{\frac{k}{m}\to \infty}       \sim \left(\frac{k}{m}\right)^s\sum_{l=0}^{n-1}\frac{(-s)_l(m-i)_l}{l!}\left(-\frac{k}{m}\right)^l,
\end{equation}
where $n$ defines the approximation quality. Combining this expansion with \eqref{eq_th-1-7} and noting that the contour integral represents Meijer G-function
yields \eqref{thm-2-2-1}. Extensive numeric simulations demonstrated that for most of the practical applications only the first 2--3 terms are enough to deliver approximation error less than 2\%. Moreover, \eqref{thm-2-2-1} can be further simplified by setting $l=0$ in Meijer G-function and evaluating the resultant  series under the assumption of $n\to\infty$, yielding \eqref{thm-2-2-2}.
\end{pf}

It should be noted that rather than applying the asymptotic series representation on the first step, as it is widely done (see, for instance, \cite{Par14}, \cite{Gar14}), it is applied on the next to last step, yielding the improved approximation. Moreover,  the derived expressions are evaluated in terms of the classical Meijer G-functions that are readily implemented in most of the modern computer algebra systems, like Matlab, Wolfram Mathematica, etc.

It can be noted that for given fading environment (i.e., fixed $\kappa$ and $\mu$), capacity improvement is mainly associated with the SNR improvement, thus, from a practical point of view, the knowledge of the capacity high-SNR behaviour is critical for the estimation of the achievable communication rate.

\begin{thm}
The ergodic capacity expression for fluctuating double-Rayleigh with line-of-sight fading channel model in case of the optimal rate adaptation strategy for high signal-to-noise ratio regime can be approximated by
\begin{equation}\label{eq_thm-4}
\bar{\bm{\mathrm{C}}}_{\text{ap}}^{\text{SNR-hi}}         \simeq        \log_2\left(\frac{\bar{\gamma }}{k+1}\right)-\frac{2\mathbb{C}_e}{\ln 2} + \frac{1}{\Gamma (m)\ln 2 }\sum_{i=0}^{\infty}\frac{H_i}{(i!)^2} G_{1,2}^{\,2,1}\!\left(\left.{\begin{matrix}1-m\\i, 1\end{matrix}}\;\right|\,\frac{k}{m}\right),
\end{equation}
where $H_i$ is the harmonic number and $\mathbb{C}_e\approx 0.577$ is the Euler-Mascheroni constant.
\end{thm}
\begin{pf}

To prove Theorem 4 one starts with the high SNR definition of ORA capacity \cite{Sil22}. Substituting \eqref{eq-pdf-conditional}, \eqref{eq-pdf-full} and applying series representation of the confluent hypergeometric function, one obtains:
\begin{eqnarray}\label{eq_th-4-1}
\bar{\bm{\mathrm{C}}}\approx\displaystyle\int_{0}^{\infty}\log_2\left(\gamma \right)f_{\gamma }(\gamma ){\rm d}\gamma = \frac{1}{\ln 2}\sum_{i=0}^{\infty}\frac{(m)_i}{(i!)^2}\int_{0}^{\infty}\frac{m^m(k+1)^{i+1}}{(k+m x)^{m+i}}\frac{x^{m-1-i}k^i}{\bar{\gamma }^{i+1}} e^{-x}\int_{0}^{\infty}\gamma^i \ln\gamma e^{-\frac{(k+1)}{x\bar{\gamma }}\gamma }{\rm d}\gamma {\rm d}x.
\end{eqnarray}

Using the fact that (see \cite{Gra07}, equation 4.352.2)
\begin{equation}\label{eq_th-4-2}
  \int_{0}^{\infty}\gamma^i \ln\gamma e^{-\frac{(k+1)}{x\bar{\gamma }}\gamma }{\rm d}\gamma=\frac{i!}{\left(\frac{k+1}{x\bar{\gamma }}\right)^{i+1}}\left(H_i-\mathbb{C}_e-\ln\left(\frac{k+1}{x\bar{\gamma }}\right) \right),
\end{equation}
the expression \eqref{eq_th-4-1} can be represented as

\begin{eqnarray}\label{eq_th-4-3}
\bar{\bm{\mathrm{C}}}\approx\displaystyle \frac{1}{\ln 2}\sum_{i=0}^{\infty}\frac{(m)_i}{i!} \int_{0}^{\infty}\frac{(m x)^m k^i}{(k+m x)^{m+i}}\left(H_i-\mathbb{C}_e-\ln\left(\frac{k+1}{x\bar{\gamma }}\right)\right) e^{-x}{\rm d}x.
\end{eqnarray}

Owing to the relation \eqref{eq_th-1-6}:
\begin{eqnarray}\label{eq_th-4-4}
\bar{\bm{\mathrm{C}}}\approx&&\displaystyle\hspace{-10pt} \frac{1}{\ln 2}\sum_{i=0}^{\infty}\frac{(m)_i}{i!}\left(\frac{k}{m}\right)^i\left(H_i-\mathbb{C}_e-\ln\left(\frac{k+1}{\bar{\gamma }}\right)\right)\Gamma (m+1)U\left(m+i, i,  \frac{k}{m}\right)\nonumber\\
&&\hspace{-10pt}+ \frac{1}{\ln 2}\sum_{i=0}^{\infty}\frac{(m)_i}{i!}\left(\frac{k}{m}\right)^i \int_{0}^{\infty}\frac{x^m}{\left(x+\frac{k}{m}\right)^{m+i}}\ln x e^{-x}{\rm d}x.
\end{eqnarray}

At this stage, it should be noted that the series in \eqref{eq_th-4-3} converges only for $\sfrac{k}{m}<1$, that implies a restriction on the further derivations. Assuming the summation in \eqref{eq_th-4-4} as an integration with the respect to the counting measure, and applying the Fubini's theorem \cite{Bil12} (which justifies the correctness of the summation and the integration interchanging), and using the facts that
\begin{eqnarray}\label{eq_th-4-5}
 &&\sum_{i=0}^{\infty}\frac{(m)_i}{i!}\left(\frac{\frac{k}{m}}{x+\frac{k}{m}}\right)^i=\frac{x^{-m}}{\left(x+\frac{k}{m}\right)^{-m}},\label{eq_th-4-5-1}  \\
&&\int_{0}^{\infty}\ln x \; e^{-x} {\rm d}x=\mathbb{C}_e,\label{eq_th-4-5-2}
\end{eqnarray}
one obtains the result for the high-SNR approximation of the ergodic capacity (under ORA policy) for the fdRLoS channel model:
\begin{eqnarray}\label{eq_th-4-6}
\bar{\bm{\mathrm{C}}}\approx&&\displaystyle\hspace{-20pt} \frac{\Gamma (m+1)}{\ln 2}\sum_{i=0}^{\infty}\frac{(m)_i}{i!}\left(\frac{k}{m}\right)^i\left(H_i-\mathbb{C}_e+\ln\left(\frac{\bar{\gamma }}{(k+1)}\right)\right)U\left(m+i, i,  \frac{k}{m}\right)-\frac{\mathbb{C}_e}{\ln 2}.
\end{eqnarray}

The result can be further simplified by using the relation between the confluent Kummer hypergeometric function and the Meijer G-function:
\begin{eqnarray}\label{eq_th-4-7}
U\left(m+i, i,  \frac{k}{m}\right)=\frac{ G_{1,2}^{\,2,1}\!\left(\left.{\begin{matrix}1-m-i\\0, 1-i\end{matrix}}\;\right|\,\frac{k}{m}\right)}{\Gamma (m+i)\Gamma (m+1)}.
\end{eqnarray}

Using \eqref{eq_th-1-4} and applying the following argument inversion (see equation 16.19.1 in \cite{DLMF}) and summation (see \cite{Pru92} equation 6.11.1.5) properties  of the Meijer G-function:
\begin{equation}\label{eq_th-4-8-0}
G_{p,q}^{\,m,n}\!\left(\left.{\begin{matrix}(a_p)\\(b_q)\end{matrix}}\;\right|\,z\right) =G_{q,p}^{\,n,m}\!\left(\left.{\begin{matrix}1-(b_q)\\1-(a_p)\end{matrix}}\;\right|\,\frac{1}{z}\right),
\end{equation}
\begin{equation}\label{eq_th-4-8-1}
\displaystyle \sum_{i=0}^{\infty}\frac{1}{i!}G_{p+1,q}^{\,m,n+1}\!\left(\left.{\begin{matrix}a-i,(a_p)\\(b_q)\end{matrix}}\;\right|\,z\right)=z^{a-1}\frac{\displaystyle \prod  _{j=1}^{m}\Gamma (b_j-a+1)\prod  _{j=1}^{n}\Gamma (a-a_j)}{\displaystyle \prod  _{j=m+1}^{q}\Gamma (a-b_j)\prod  _{j=n+1}^{p}\Gamma (a_j-a+1)},
\end{equation}
helps to state that
\begin{eqnarray}\label{eq_th-4-8}
\sum_{i=0}^{\infty}\frac{ G_{1,2}^{\,2,1}\!\left(\left.{\begin{matrix}1-m\\i, 1\end{matrix}}\;\right|\,\frac{k}{m}\right)}{i!}=\Gamma (m).
\end{eqnarray}

Assuming \eqref{eq_th-4-8} and reorganizing the terms in \eqref{eq_th-4-6}, one obtains \eqref{eq_thm-4}, finalizing the proof.
\end{pf}

\subsection{OPRA case}
\begin{thm}
The ergodic capacity expression for fluctuating double-Rayleigh with line-of-sight fading channel model in the case of the optimal power and rate adaptation strategy can be represented as
\begin{eqnarray}\label{eq_thm-5}
\bar{\bm{\mathrm{C}}}=\frac{k}{ \Gamma (m+1)\ln 2 }\sum_{i=0}^{\infty}\frac{G_{1,1:3,2:0,1}^{1,1:0,3:1,0}\left(
\hspace{-5pt} \begin{array}{c}
-m\\
1+i\\
\end{array}\hspace{-2pt}\middle\vert
\begin{array}{c}
-i, 1, 1\\
0, 0\\
\end{array}\hspace{-2pt}\middle\vert
\begin{array}{c}
\mbox{---} \\
0\\
\end{array}\middle\vert
\frac{\bar{\gamma }k}{\gamma_0(k+1)}, \frac{k}{m}
\right)}{(i!)^2}.
\end{eqnarray}
\end{thm}
\begin{pf}
To prove Theorem 5 one starts with the change of variables $\gamma=\gamma_0 t$ in initial definition \eqref{eq-cap-def-1-OPRA}:
\begin{eqnarray}
 \bar{\bm{\mathrm{C}}}\!\!\!\!&=&\!\!\!\!\displaystyle\int_{\gamma_0}^{\infty}\log_2\left(\frac{\gamma }{\gamma_0} \right)f_{\gamma }(\gamma ){\rm d}\gamma =\frac{\gamma_0}{\ln 2}\displaystyle\int_{1}^{\infty}\ln t\;f_{\gamma }(\gamma_0 t){\rm d}t \label{eq_th-5-1}\\
 &=&\!\!\!\frac{\gamma_0}{\ln 2}\displaystyle\int_{1}^{\infty}\frac{m^m(k+1)}{(k+m x)^m}\frac{x^{m-1}}{\bar{\gamma }}\;e^{-x}\sum_{i=0}^{\infty}\frac{(m)_i}{(i!)^2}\left(\frac{k(k+1)}{x(k+m x)\bar{\gamma }}\gamma_0\right)^i \int_{1}^{\infty}\ln t\;t^i\;e^{-\frac{(k+1)}{x\bar{\gamma }}\gamma_0 t}{\rm d}t {\rm d}x, \label{eq_th-5-2}
\end{eqnarray}
The inner integral in \eqref{eq_th-5-2} can be evaluated in the following manner:
\begin{eqnarray}
\displaystyle \int_{1}^{\infty}\ln t\;t^i\;f_{\gamma }(\gamma_0 t)e^{-\frac{(k+1)}{x\bar{\gamma }}\gamma_0 t}{\rm d}t\!\!\!\!&=&\!\!\!\!\frac{\partial}{\partial a}\left[ \int_{1}^{\infty}\;t^{a+i}e^{-\frac{(k+1)}{x\bar{\gamma }}\gamma_0 t}{\rm d}t \right]\bigg|_{a\to 0} \label{eq_th-5-3}\\
&=&\!\!\!\!\frac{\partial}{\partial a}\left[ E_{-(a+i)}\left( \frac{(k+1)}{x\bar{\gamma }}\gamma_0\right) \right]\bigg|_{a\to 0}\label{eq_th-5-4} \\
&=&\!\!\!\!\frac{ G_{2,3}^{\,3,0}\!\left(\left.{\begin{matrix}1-i, 1-i\\1, -i, -i\end{matrix}}\;\right|\,\frac{(k+1)}{x\bar{\gamma }}\gamma_0\right)}{\frac{(k+1)}{x\bar{\gamma }}\gamma_0},\label{eq_th-5-5}
\end{eqnarray}
where $E_{n}(z)$ is generalized exponential integral \cite{DLMF}.

Applying the argument inversion property of the Meijer G-function (see equality \eqref{eq_th-4-8-0}), reorganizing the terms, and applying the identity \eqref{eq_th-1-4} yield:
 \begin{equation}\label{eq_th-5-7}
\bar{\bm{\mathrm{C}}}=\displaystyle\frac{m^m}{\ln 2}\sum_{i=0}^{\infty}\frac{(m)_i}{(i!)^2}\left(\frac{\gamma_0k(k+1)}{\bar{\gamma }}\right)^i
\int_{0}^{\infty} x^{m-i}(k+m x)^{-m-i}
G_{3,2}^{\,0,3}\!\left(\left.{\begin{matrix}0, i+1, i+1\\i, i\end{matrix}}\;\right|\,\frac{\bar{\gamma }x}{\gamma_0(k+1)}\right)e^{-x}  {\rm d}x.
\end{equation}

Sequentially applying the change of variable $z=\frac{\bar{\gamma }x}{\gamma_0(k+1)}$, the transformation \eqref{eq_lem-1-2} for the exponential term in \eqref{eq_th-5-7}, and the following connection:
\begin{equation}\label{eq_th-5-8}
\left(z+\frac{k\bar{\gamma }}{\gamma_0(k+1)m}\right)^{-i-m}=\left(\frac{k\bar{\gamma }}{\gamma_0(k+1)m}\right)^{-i-m}\frac{G_{1,1}^{\,1,1}\!\left(\left.{\begin{matrix}1-m-i\\0\end{matrix}}\;\right|\,\frac{\gamma_0(k+1)m}{k\bar{\gamma }}z\right)}{\Gamma (m+i)}
\end{equation}
yields the representation for the OPRA channel capacity:
\begin{eqnarray}\label{eq_th-5-9}
\bar{\bm{\mathrm{C}}}=\left(\frac{k\bar{\gamma }}{\gamma_0(k+1)m}\right)^{-1-m}\frac{k}{\Gamma (m+1)\ln 2}\sum_{i=0}^{\infty}\frac{1}{(i!)^2}\hspace{-15pt}&&\int_{0}^{\infty}G_{1,1}^{\,1,1}\!\left(\left.{\begin{matrix}1-m-i\\0\end{matrix}}\;\right|\,\frac{\gamma_0(k+1)m}{k\bar{\gamma }}z\right)       \times       \nonumber\\ &&       \times        G_{3,2}^{\,0,3}\!\left(\left.{\begin{matrix}-i, 1, 1\\0, 0\end{matrix}}\;\right|\,z\right)G_{0,1}^{\,1,0}\!\left(\left.{\begin{matrix}\mbox{---}\\0\end{matrix}}\;\right|\,\frac{\gamma_0(k+1)\gamma }{z\bar{\gamma }}\right)
{\rm d}z.
\end{eqnarray}

Making use of the Meijer G-function integration theorem (see \cite{Wol21}) finalizes the proof.
\end{pf}

It should be specifically pointed out that the derived expression is valid for arbitrary positive values of the shadowing parameter $m$. Although the result is given in terms of the infinite series, as it was already mentioned above, only several terms are needed for reasonable accuracy that helps to obtain a computational speedup compared to the direct numerical integration in \eqref{eq-cap-def-1-OPRA}. Moreover, a further series convergence improvement can be obtained via Shanks transformation or Richardson extrapolation \cite{Ben99} used herein for further numerical analysis.

Similarly to the ORA case, the capacity analysis is  completed by deriving the high SNR asymptotics of the expression \eqref{eq_thm-5}.
\begin{thm}
The ergodic capacity expression for fluctuating double-Rayleigh with line-of-sight fading channel model in case of the optimal power and rate adaptation strategy for high signal-to-noise ratio regime can be approximated by
\begin{equation}\label{eq_thm-6}
\bar{\bm{\mathrm{C}}}_{\text{ap}}^{\text{SNR-hi}}         \simeq        \log_2\left(\frac{\bar{\gamma }}{\gamma_0(k+1)}\right)-\frac{\mathbb{C}_e}{\ln 2} + \frac{1}{\Gamma (m)\ln 2 }\sum_{i=0}^{\infty}\frac{\psi^{(0)}(i+1)}{i!} G_{1,2}^{\,2,1}\!\left(\left.{\begin{matrix}1-m\\i, 1\end{matrix}}\;\right|\,\frac{k}{m}\right),
\end{equation}
where $\psi^{(0)}(i+1)$ is the digamma function \cite{DLMF}.
\end{thm}
\begin{pf}
To prove Theorem 6 one starts with the expression \eqref{eq_th-5-7} and expand Meijer G-function in case of $\bar{\gamma }\to \infty$:
\begin{equation}\label{eq_th-6-1}
  G_{3,2}^{\,0,3}\!\left(\left.{\begin{matrix}0, i+1, i+1\\i, i\end{matrix}}\;\right|\,\frac{\bar{\gamma }x}{\gamma_0(k+1)}\right)\bigg|_{\bar{\gamma }\to \infty}\approx \left(\frac{\bar{\gamma }x}{\gamma_0(k+1)}\right)^i\ln\left(\frac{\bar{\gamma }x}{\gamma_0(k+1)}\right) i!+\left(\frac{\bar{\gamma }x}{\gamma_0(k+1)}\right)^i i!\psi^{(0)}(i+1).
\end{equation}

Combining \eqref{eq_th-5-7} and \eqref{eq_th-6-1} yields:
\begin{eqnarray}\label{eq_th-6-2}
 \bar{\bm{\mathrm{C}}}_{\text{ap}}^{\text{SNR-hi}}\approx \displaystyle\frac{1}{\ln 2}\sum_{i=0}^{\infty}\frac{(m)_i}{i!}\left(\frac{k}{m}\right)^i \!\!\! \!\!\!&&\!\!\! \!\!\! \left[\int_{0}^{\infty} x^{m}\left(\frac{k}{m}+ x\right)^{-m-i}\ln\left(\frac{\bar{\gamma }x}{\gamma_0(k+1)}\right) e^{-x}  {\rm d}x +\right.\nonumber\\
&+&\!\!\! \left.\psi^{(0)}(i+1)\int_{0}^{\infty} x^{m}\left(\frac{k}{m}+ x\right)^{-m-i}e^{-x}  {\rm d}x \right].
\end{eqnarray}

For the first part of the right hand side of the equation \eqref{eq_th-6-2} one can switch integration and summation (by virtue of the Fubini's theorem), and, verifying the convergence, with the help of \eqref{eq_th-4-5-1}-\eqref{eq_th-4-5-2} express it as:
\begin{equation}\label{eq_th-6-3}
\displaystyle\frac{1}{\ln 2}\sum_{i=0}^{\infty}\frac{(m)_i}{i!}\left(\frac{k}{m}\right)^i \int_{0}^{\infty} x^{m}\left(\frac{k}{m}+x\right)^{-m-i}{\rm d}x=\log_2\left(\frac{\bar{\gamma }}{\gamma_0(k+1)}\right)-\frac{\mathbb{C}_e}{\ln 2}.
\end{equation}

For the second term of the right hand side of the equation \eqref{eq_th-6-2} one can use the integral representation of the confluent Kummer hypergeometric function \eqref{eq_th-1-6}. Performing the transformation \eqref{eq_th-4-7} and simplifying the expression finally yield:
 \begin{equation}\label{eq_th-6-4}
 \bar{\bm{\mathrm{C}}}_{\text{ap}}^{\text{SNR-hi}}         \simeq        \log_2\left(\frac{\bar{\gamma }}{\gamma_0(k+1)}\right)-\frac{\mathbb{C}_e}{\ln 2} + \frac{1}{\Gamma (m)\ln 2 }\sum_{i=0}^{\infty}\frac{\psi^{(0)}(i+1)}{i!} G_{1,2}^{\,2,1}\!\left(\left.{\begin{matrix}1-m\\i, 1\end{matrix}}\;\right|\,\frac{k}{m}\right),
 \end{equation}
 which justifies the claim of Theorem 6.
\end{pf}

\begin{cor}
  The high-SNR ergodic capacity approximations for the fdRLoS channel in cases of ORA and OPRA policies are equal up to the change of variable $\bar{\gamma }^{\mathrm{ORA}}=\bar{\gamma }^{\mathrm{OPRA}}\gamma_0$.
\end{cor}

\begin{pf}
  The derived expressions for ORA \eqref{eq_thm-4} and OPRA \eqref{eq_thm-6} high-SNR capacity approximations can be easily connected by noting the relation between the digamma function and harmonic number, i.e., $\psi^{(0)}(i+1)=H_i-\mathbb{C}_e$ (see equation $5.4.14$ in \cite{DLMF}) and using the obtained result \eqref{eq_th-4-8}.

  It can be pointed out that for an the increasing average SNR the threshold value $\gamma_0\to 1$ thus the expressions for ORA and OPRA capacities coincide.
\end{pf}

To the best of the author's knowledge, the capacity analysis of the fdRLoS channel is absent in current technical literature, and the results \eqref{lem-1}, \eqref{thm-1}, \eqref{thm-2-1}-\eqref{thm-2-2-2}, \eqref{eq_thm-4}, \eqref{eq_thm-5}, \eqref{eq_thm-6} are novel and have not been reported previously.

\section{Simulation and results}
To verify the correctness of the derived closed-form solutions (see Theorems 2 and 5) and approximations (see Theorems 3, 4 and 6) numeric simulation was performed.

\begin{figure}[!t]
\vspace{-5pt}\centerline{\includegraphics[width=\columnwidth]{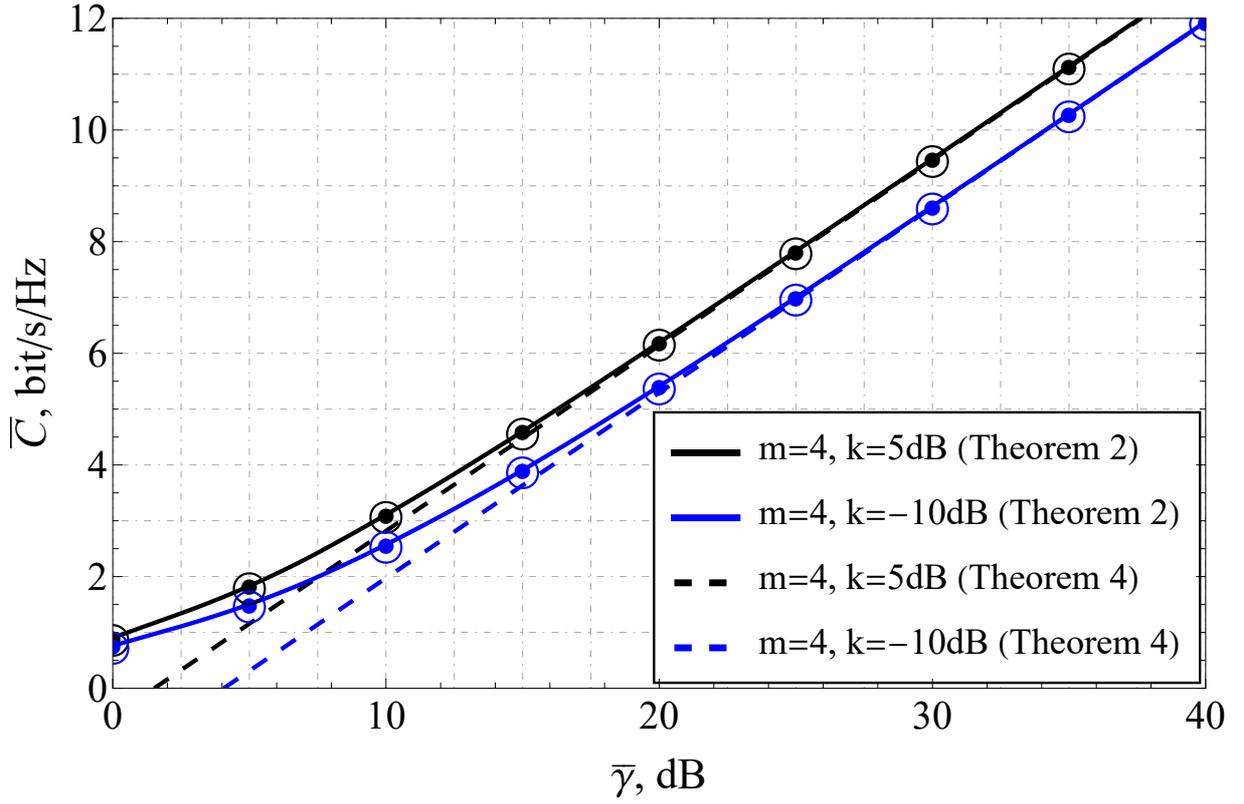}}
\caption{Ergodic capacity per unit bandwidth of the fdRLoS fading channel for ORA scheme}
\label{fig1}
\end{figure}

\begin{figure}[!h]
\vspace{-5pt}\centerline{\includegraphics[width=\columnwidth]{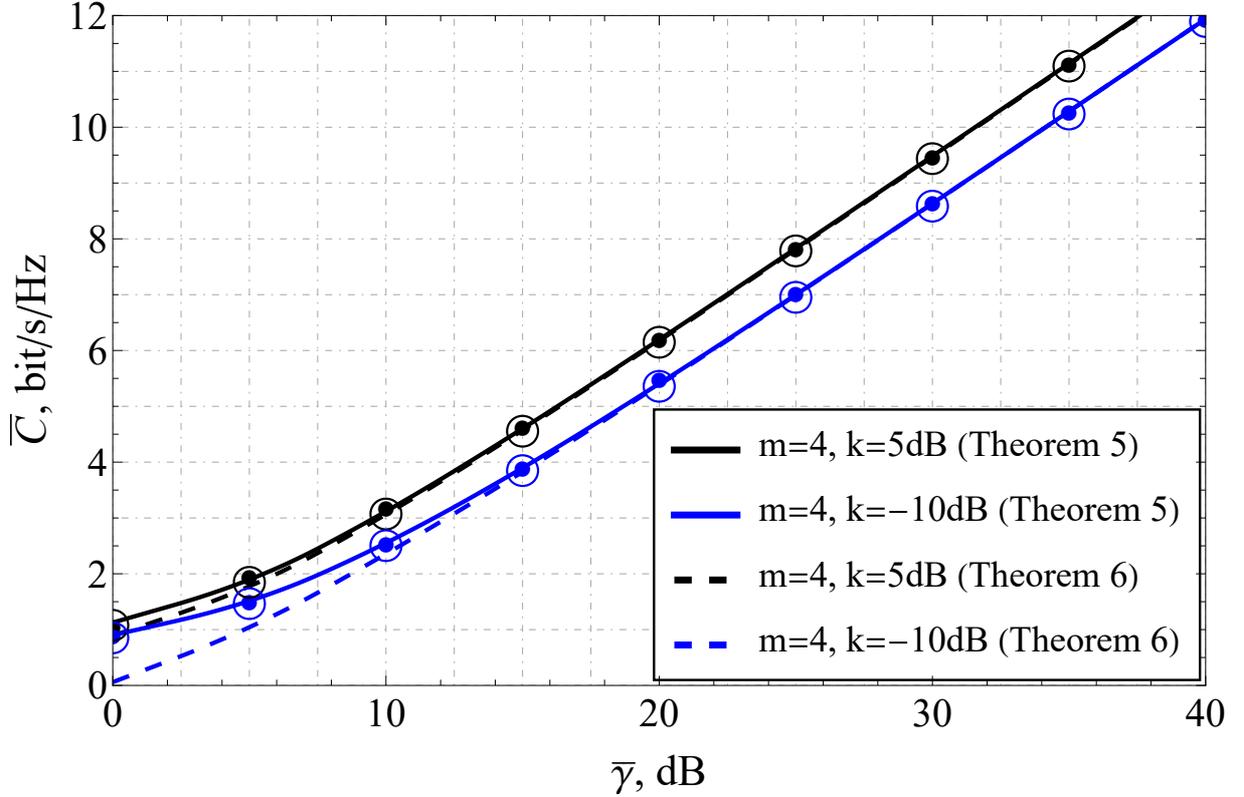}}
\caption{Ergodic capacity per unit bandwidth of the fdRLoS fading channel for OPRA scheme}
\label{fig2}
\end{figure}

\subsection{Capacity analysis}
The results for the average SNR dependence of the ergodic capacity are presented in Fig~(\ref{fig1}) for ORA and Fig~(\ref{fig2}) for OPRA. For both plots, the notation is as follows:
\begin{itemize}
\item the solid lines are obtained via the derived closed-form solutions \eqref{thm-1} (for ORA) and \eqref{eq_thm-5} (for OPRA);
\item the dashed lines represent the derived high-SNR approximations \eqref{eq_thm-4} (for ORA) and \eqref{eq_thm-6} (for OPRA) respectively;
\item the filled markers are obtained with direct numerical integration with \eqref{eq-pdf-full}, \eqref{eq-pdf-conditional}  substituted in to \eqref{eq-cap-def-1-ORA} (for ORA) and \eqref{eq-cap-def-1-OPRA}  (for OPRA);
\item the empty circled-markers are obtained with Monte-Carlo simulation (averaged over $10^8$ samples).
\end{itemize}

The results clearly show that the derived expressions perfectly match numerical solution (derived via brute-force integration) and simulation. Moreover, the proposed approximations are tight enough for not very high values of the average SNR, i.e., for $\bar{\gamma }\geq 15\div 20$~dB in case of ORA and $\bar{\gamma }\geq 7\div 10$~dB in case of OPRA. The approximation improves with the increase in $k$. It was determined that the change of $m$ does not impact $\bar{\bm{\mathrm{C}}}$ sufficiently (as will be demonstrated further).

It can be observed that the results for OPRA are very close (within several percent) to the ones for ORA with the sight advantage in the case of low SNR (as expected from Corollary 6.1). Thus, for further analysis, only the ORA case will be considered.

\begin{figure}[!t]
\vspace{-5pt}\centerline{\includegraphics[width=\columnwidth]{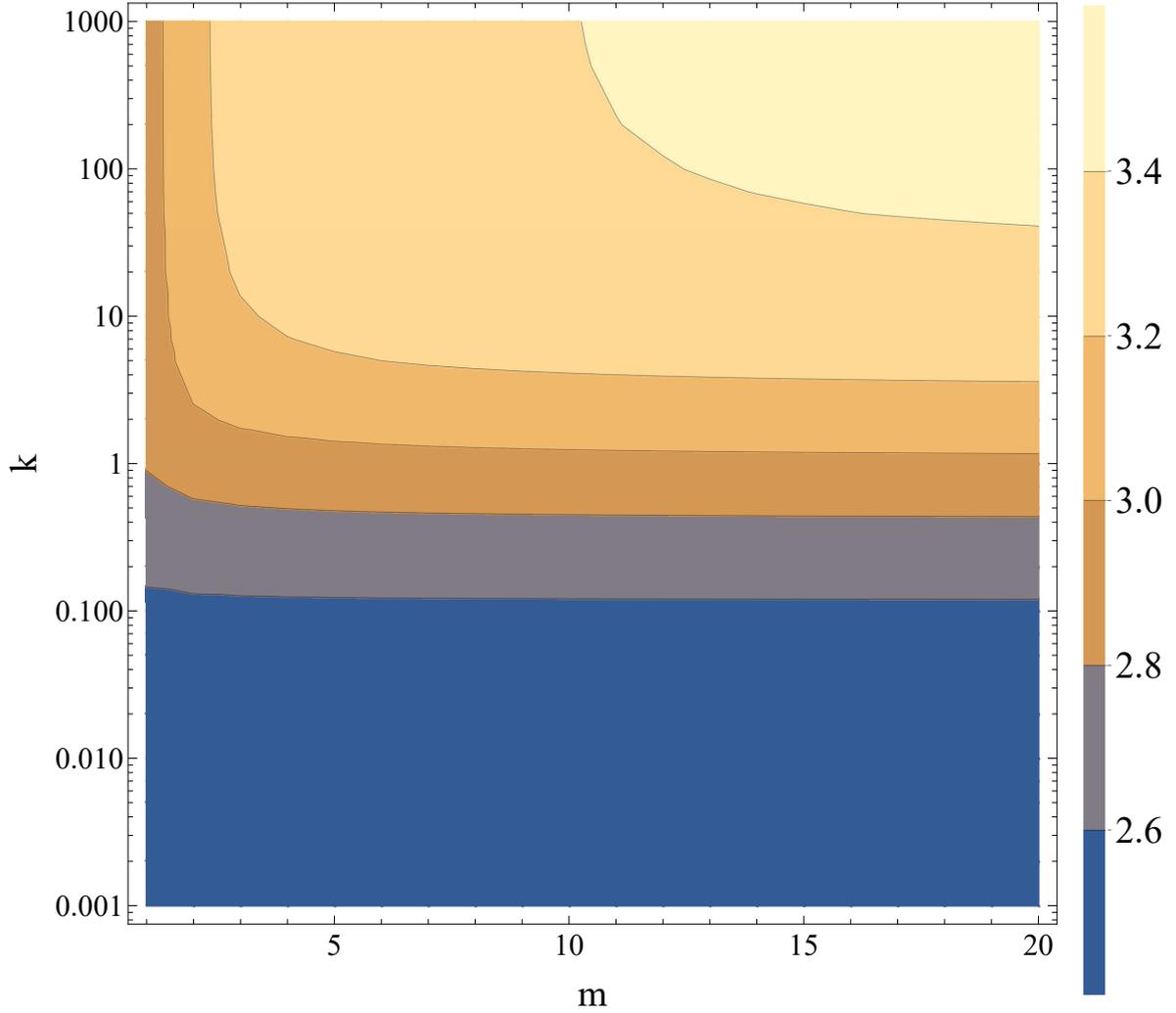}}
\caption{Ergodic capacity per unit bandwidth of the fdRLoS fading channel for ORA scheme with variable $m$ and $k$ and $\bar{\gamma }=10$~dB}
\label{fig3}
\end{figure}

To study the joint impacts of $m$ and $k$ on the ergodic capacity, a contour map was plotted (see Fig~(\ref{fig3})). It was obtained with the combination of the derived solutions for $m     \notin      \mathbb{Z}^{+}_0$ and $m     \in \mathbb{Z}^{+}_0$ to cover all the possible cases for $m\geq 0.5$.  It was observed that for all possible range of parameters $2.4$~bit/s/Hz$\leq\bar{\bm{\mathrm{C}}}\leq 3.5$~bit/s/Hz.
In real-life applications the channel parameters are usually unknown and should be estimated in the process from the available measurements, and the estimation accuracy impacts the overall performance. This means that it is necessary to understand the effect of the estimated parameters fluctuation on the system capacity. For instance, if the impact is negligible (i.e., an asymptotic regime in the corresponding parameter), coarse yet fast inference methods are preferable; elsewise, more complex algorithms (which are generally slower) are required.

Analyzing the obtained results, one can notice that in the region with low $k$ the impact of the shadowing parameter's variation is negligible (as it was observed in Fig~(\ref{fig1}) and Fig~(\ref{fig2)}); therefore, even $m=1$ can be assumed as asymptotic. The situation changes for the high Rician factor: the increase of $m$ delivers some gain in $\bar{\bm{\mathrm{C}}}$: about $0.5$~bit/s/Hz. Moreover, for reasonable values of the Rician factor (i.e. $k\leq 30$~dB) the sufficient change appears for $m\leq 4$.

\subsection{Proposed approximation analysis}

As it was already mentioned, the derived expressions for the high-SRN approximation demonstrate good correspondence with the closed-form results even in the case of moderate values of $\bar{\gamma }$. To estimate the quality of $\sfrac{\kappa }{\mu }$ approximations (see Theorem 3), numerical simulation was performed.

\begin{figure}[!t]
\centerline{\includegraphics[width=\columnwidth]{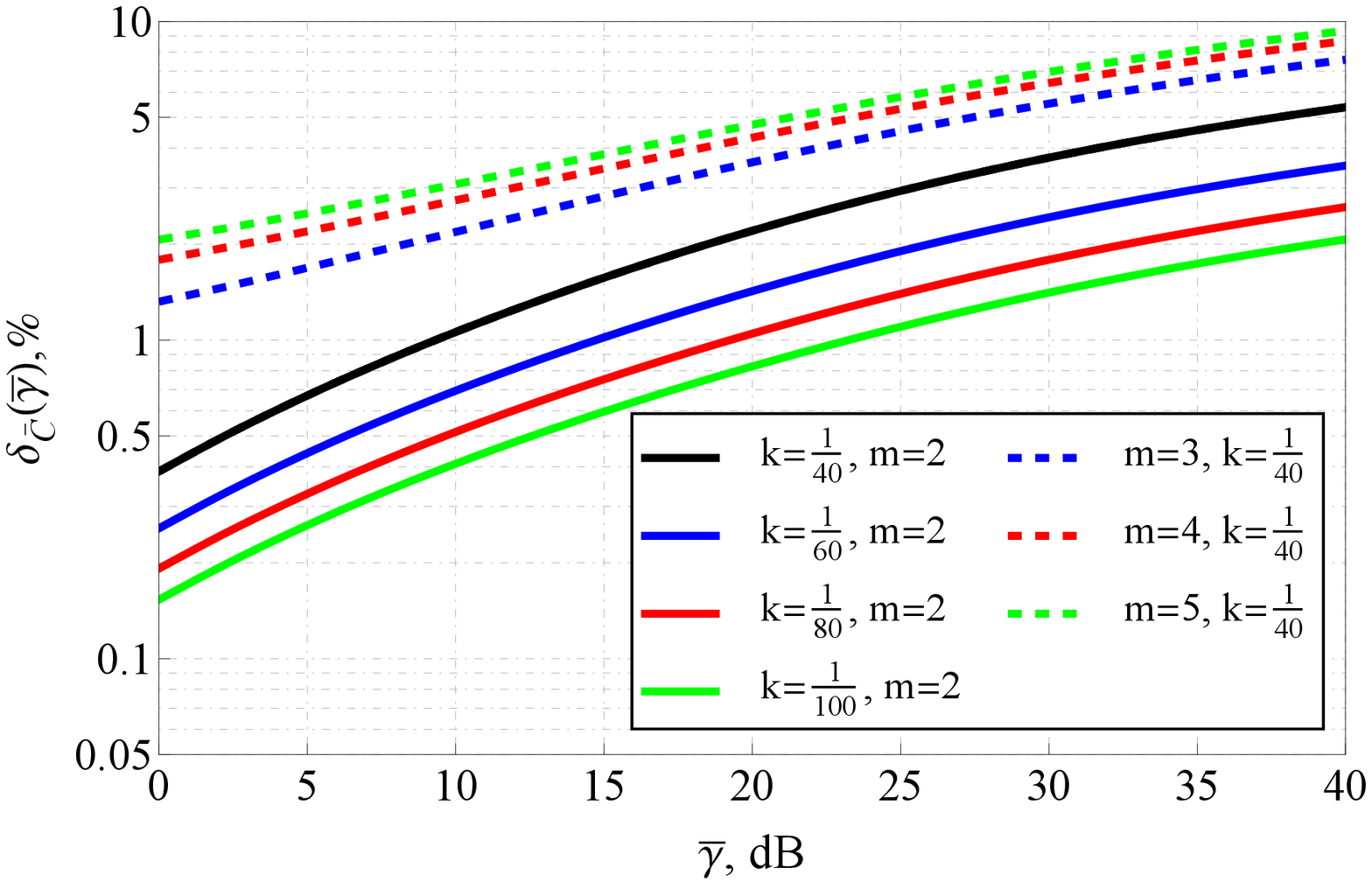}}
\caption{Ergodic capacity approximation error of the fdRLoS fading channel  for ORA scheme and $\sfrac{k}{m}\to 0$}
\label{fig4}
\end{figure}

\begin{figure}[!h]
\centerline{\includegraphics[width=\columnwidth]{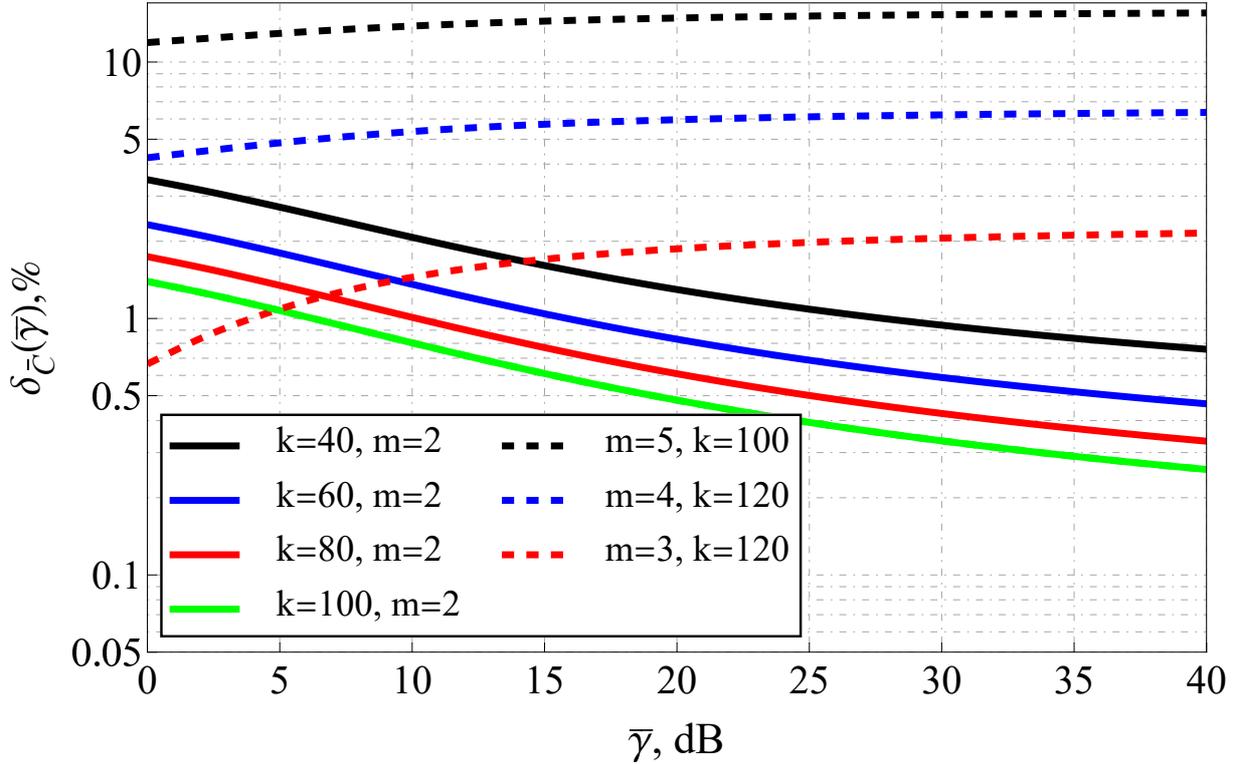}}
\caption{Ergodic capacity approximation error of the fdRLoS fading channel  for ORA scheme and $\sfrac{k}{m}\to\infty$}
\label{fig5}
\end{figure}

Table I exemplifies the comparison of the results for the case $\sfrac{k}{m}\to \infty$, $m=2$ and variable $\bar{\gamma }$. The results for $\bar{\bm{\mathrm{C}}}_{\text{num}}$ were derived via numerical integration in \eqref{eq-cap-def-1-ORA} with \eqref{eq-pdf-conditional} and  \eqref{eq-pdf-full} being substituted,  $\bar{\bm{\mathrm{C}}}_{\text{th}}$ was derived via \eqref{thm-1} and algorithm, proposed in \cite{Gar14}, and $\bar{\bm{\mathrm{C}}}_{\text{ap}}^{\text{hi}}$ was evaluated via the simplified approximation \eqref{thm-2-2-2}.
The obtained results demonstrated the complete agreement between the derived solution and numerical integration (up to the computational accuracy of the applied-for numeric integration algorithms) and high correspondence with the proposed approximation. Since the difference is relatively small for further analysis, it is appropriate to study the relative error (see Fig~(\ref{fig4})-(\ref{fig5})) defined as $\delta_{\bar{\bm{\mathrm{C}}}}=\left|\frac{\bar{\bm{\mathrm{C}}}-\bar{\bm{\mathrm{C}}}_{\text{ap}}}{\bar{\bm{\mathrm{C}}}}\right|$, where
\begin{itemize}
  \item for plots in Fig~(\ref{fig4}) $\bar{\bm{\mathrm{C}}}_{\text{ap}}=\bar{\bm{\mathrm{C}}}_{\text{ap}}^{\text{low}}$, and evaluated with approximation \eqref{thm-2-1};
  \item for plots in Fig~(\ref{fig5}) $\bar{\bm{\mathrm{C}}}_{\text{ap}}=\bar{\bm{\mathrm{C}}}_{\text{ap}2}^{\text{hi}}$, and evaluated via expression \eqref{thm-2-2-2}.
\end{itemize}

Moreover, since one can achieve  $\sfrac{k}{m}\to 0$ and  $\sfrac{k}{m}\to \infty$ in two ways: increasing/decreasing the numerator, keeping a fixed denominator, and vice versa, both cases were plotted, but in such a way to maintain the same ratio $\sfrac{k}{m}$.

\begin{table}[!t]
\caption{The ergodic capacity of the fluctuating double-Rayleigh with line-of-sight fading channel for  ORA scheme}
\label{table_example}
\centering
\begin{tabular}{|c|c|c|c||c|c|c|}
\addlinespace \hline
$\bar{\gamma }$,~dB&\makecell{$\bar{\mathbf{C}}_{\text{th}}$\\ $ \kappa=20$} &\makecell{$\bar{\mathbf{C}}_{\text{num}}$\\ $ \kappa=20$}  & \makecell{$\bar{\mathbf{C}}_{\text{ap}}^{\text{hi}}$\\ $ \kappa=20$}&\makecell{$\bar{\mathbf{C}}_{\text{th}}$\\ $ \kappa=200$} &\makecell{$\bar{\mathbf{C}}_{\text{num}}$\\ $ \kappa=200$}  & \makecell{$\bar{\mathbf{C}}_{\text{ap}}^{\text{hi}}$\\ $ \kappa=200$}\\
\hline
0& 0.91 & 0.91 & 0.94 &0.92  & 0.92 & 0.92 \\
\hline
10& 3.13  & 3.13 & 3.32 & 3.16 & 3.16 & 3.18\\
\hline
20& 6.22 & 6.22 & 6.70 & 6.27 & 6.27 & 6.32\\
\hline
30& 9.56 & 9.56 & 10.32 &9.62  & 9.62 & 9.65\\
\hline
40& 12.84 & 12.84 & 13.97 &12.89  & 12.89 & 13.01\\
\hline
\end{tabular}
\end{table}

It can be noted (see Fig~(\ref{fig1})) that for most of the practically meaningful average SNRs (no exceeding $25\div30$~dB) the relative error for
$\bar{\bm{\mathrm{C}}}_{\text{ap}}^{\text{low}}$ does not exceed $5\div10\%$, for small $m$ (i.e. $m=2$) in combination with $k\leq \sfrac{1}{100}$, $\delta_{\bar{\bm{\mathrm{C}}}}\leq 1.5\%$ for all of the assumed $\bar{\gamma }$.
It should be noted that the approximation quality improves with the decrease of $k$ and degrades with an increase of $m$. That means that it can be efficiently used in cases of high-to-moderate shadowing
and small Rician k-factor (i.e. $-16$~dB or smaller, as in Fig~(\ref{fig1})).

For the high $\sfrac{k}{m}$ approximation (see Fig~(\ref{fig1})), the overall dependence from the parameters is the same, but with a sufficient difference in behavior depending on $\bar{\gamma }$: for the increasing $m$, the relative error of the proposed approximation becomes almost insensitive to $\bar{\gamma }$, whereas the increase of $k$ (with fixed $m$) actually improves $\delta_{\bar{\bm{\mathrm{C}}}}$ with SNR. Moreover, it should be noted that $\delta_{\bar{\bm{\mathrm{C}}}}\leq 2\%$ (see solid lines), even when the 1-term approximation was used (i.e., $n=1$ in \eqref{thm-2-2-1}). Increasing the number of terms, for example, to $n=3$ helps to even more increase the approximation quality, i.e., $\delta_{\bar{\bm{\mathrm{C}}}}<0.5\%$.

\section{Conclusions}
The research presents the derived results for the ergodic capacity of a communication system in the presence of fading described by the fluctuating double-Rayleigh with line-of-sight channel model for the cases of the optimal rate adaptation strategy with constant transmit power and the optimal simultaneous power and rate adaptation strategy. For the ORA case with arbitrary noninteger fading parameter $m$, a closed-form expression was derived for the conditional capacity, which can be further used to evaluate the ergodic capacity, and for the case of the integer fading parameter, a closed-form expression for the ergodic capacity was derived in terms of the extended generalized bivariate Meijer G-function. To facilitate the computations, approximating expressions are deduced for the high average signal-to-noise ratio scenario, the cases of $\sfrac{k}{m}\to 0$ and $\sfrac{k}{m}\to \infty$. A connection between the ORA and OPRA ergodic capacity expressions in the case of high signal-to-noise ratio is established. All the derived expressions are verified by comparing with the brute-force numerical integration and simulation and demonstrated high correspondence with one another.





\end{document}